\documentclass[aps,prd,reprint,superscriptaddress,preprintnumbers,nofootinbib]{revtex4-1}

\usepackage{amsmath}
\usepackage{amsfonts} 
\usepackage{amssymb}
\usepackage{graphicx}
\usepackage{hyperref}
\usepackage{tensor}
\usepackage{siunitx}
\newcommand{\be}{\begin{equation}}
\newcommand{\ee}{\end{equation}}

\begin{document}

\preprint{MPP-2019-135, LMU-ASC 28/19}
\preprint{}

\title{\Large A Note on String Excitations and the Higuchi Bound}

             \author{\large Dieter L\"ust}
\affiliation{Arnold-Sommerfeld-Center for Theoretical Physics, Ludwig-Maximilians-Universit\"at, 80333 M\"unchen, Germany}
\affiliation{Max-Planck-Institut f\"ur Physik (Werner-Heisenberg-Institut),\\
           F\"ohringer Ring 6,
             80805, M\"unchen, Germany}
\author{\large Eran Palti}
\affiliation{Max-Planck-Institut f\"ur Physik (Werner-Heisenberg-Institut),\\
             F\"ohringer Ring 6,
             80805, M\"unchen, Germany}

\begin{abstract}
\vspace{0.5cm}
In this brief note we consider the interaction between high spin excitations in string theory along the Regge trajectory and the Higuchi bound in de Sitter space. There is always a point along the Regge trajectory where the Higuchi bound is violated. However, this point precisely coincides with a string whose length is of order the de Sitter Hubble scale. String theory therefore manifests no immediate inconsistency as long as the string scale $M_s$ is above the Hubble scale $H$. However, an implication is that the Regge trajectory must be significantly modified at some ultraviolet scale. Insisting that this modification should occur no earlier than the Planck scale would lead to a bound on the string scale of $M_s > \sqrt{H M_p}$. 
\vspace{1cm}
\end{abstract}

\maketitle

\section{Introduction}
Understanding whether de Sitter space can be consistently realised in string theory is a long standing problem. Recently, this topic gained impetus due to a conjecture which implies that de Sitter vacua are not allowed in string theory \cite{Obied:2018sgi}. We refer to \cite{Palti:2019pca} for a review. In this brief note we consider the Higuchi bound as a consistency condition for string theory in de Sitter space. Overall, we find that there is no apparent inconsistency as long the string scale $M_s$ is larger than the Hubble scale $H$, which is the expected condition. The note is to expose a few details of this simple but interesting interaction. We also comment on the possibility that if we require the Regge trajectory of string theory to remain unmodified from its flat-space behaviour up to the Planck scale, then one finds the stronger condition $M_s > \sqrt{H M_p}$. Such a condition on the Regge trajectory would be necessary for the usual flat-space softening of ultraviolet divergences to remain unmodified, but it is not clear that this needs to be the case in de Sitter space. 

\section{Regge trajectory and Higuchi bound}

The Higuchi bound \cite{Higuchi:1986py} is a bound on the mass of fields with spin two or higher in de Sitter space which captures the absence of ghosts. It states that a field of spin $l$, in $d$-dimensional de Sitter space, has a bound on its mass $M_{(l)}$ given by
\be
M^2_{(l)} \geq H^2 \left(l-1 \right) \left(d+l-4 \right) \;.
\label{hb}
\ee
The bound can be understood as the requirement that the pseudo-scalar mode of the field has positive kinetic terms and so does not yield a ghost.\footnote{More generally, for the mode of helicity $t$ there is a bound $M^2_{(l,t)} \geq H^2 \left(l-t-1 \right) \left(d+l+t-4 \right) $. When the bound is saturated the field is said to be partially massless to depth $t$.} 

The Higuchi bound is exact for de Sitter space, but is also expected to hold as long as the de Sitter isometries are only slightly broken. It therefore is expected to also hold approximately for inflation models, see for example \cite{Arkani-Hamed:2015bza,Lee:2016vti}. 

String theory intrinsically and inevitably contains massive higher spin states in the form of oscillator modes of the string. The Higuchi bound is therefore a universal bound on the Hubble scale within any perturbative string theoretic setting. The most straightforward application is to the first massive higher-spin oscillator mode, so the spin-2 state. This state has a mass $M^2_{(2)}=2M^2_s$, where $M_s$ is the string scale, and therefore bounds the string scale in de Sitter space
\be
M^2_s \geq \frac{d-2}{2} H^2 \;.
\label{stconsp2}
\ee
This is similar to an effective field theory constraint which ensures that the string oscillator modes may be neglected $M_s \gg H$. See, for example, \cite{Baumann:2014nda,Arkani-Hamed:2015bza}. However, it is important to state that (\ref{stconsp2}) is an intrinsic consistency constraint which, given that string theory contains no ghosts, cannot be violated in any valid string construction.\footnote{This is subject to the discussion in section \ref{sec:req}.}\footnote{It is natural to expect that continuously approaching the saturation of the Higuchi bound, or more generally partial masslessness, should occur at infinite distance in field space and be accompanied by an infinite tower of states becoming massless. This would be consistent with the Spin-2 conjecture \cite{Klaewer:2018yxi}.}

\begin{figure}
\centering
 \includegraphics[width=0.45\textwidth]{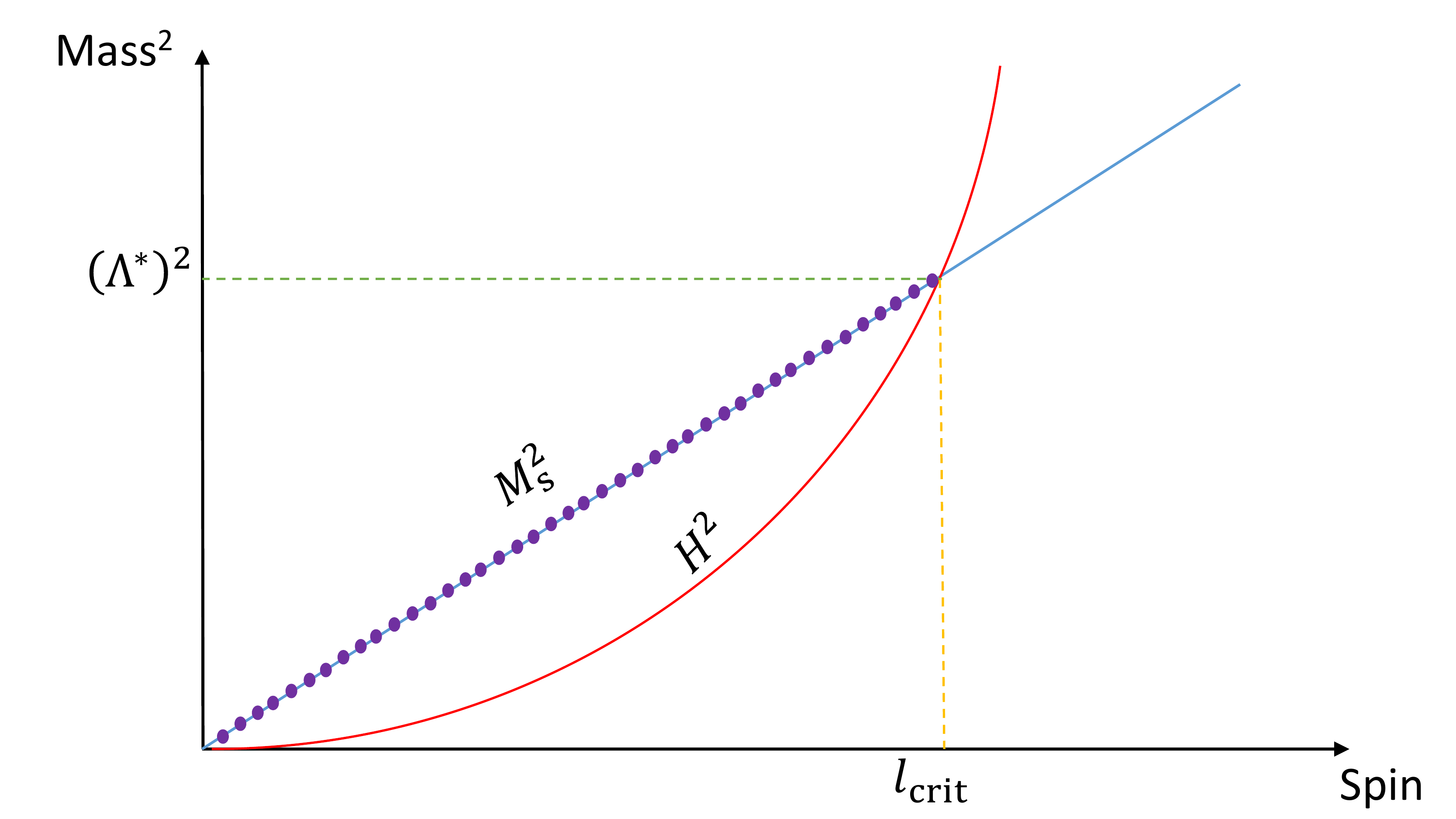}
\caption{Figure illustrating the two trajectories of mass-squared against spin. The quadratic red line is the Higuchi bound, while the linear blue line is the string Regge trajectory. At the crossing point the Higuchi bound is violated, and so if the perturbative string is to be consistent on such a background the Higuchi analysis must break down at a scale $\Lambda^*$. }
\label{fig:sw}
\end{figure}

A primary focus of this note is to study the implications not of the first higher spin string mode, but of the whole tower of increasing spin states. This is particularly interesting because for large spin $l \gg 1$ the Higuchi bound (\ref{hb}) scales quadratically with the spin $M^2_{(l)} \gtrsim H^2 l^2$. However, the oscillator modes of the string follow instead the Regge trajectory which scales only {\it linearly} with spin 
\be
M^2_{(l)} = M_s^2 l \;.
\label{rt}
\ee
Therefore, for some spin $l=l_{\mathrm{crit}}$ the bound will be violated. The critical spin is at
\be
l_{\mathrm{crit}} \sim  \left(\frac{M_{\mathrm{s}}}{H}\right)^2\;, \;\; M^2_{l_{\mathrm{crit}}} \sim \frac{M_s^4}{H^2}  \;.
\label{critspin}
\ee
If we trusted the application of the Higuchi bound to the oscillator spectrum of string theory, as derived in flat space, to arbitrarily high spins, perturbative string theory would be inconsistent in de Sitter space for any finite value of $H$. However, we can impose a cutoff $\Lambda^*$ on the scale at which the Regge trajectory, or more precisely the application of the Higuchi bound to oscillator states, must break down. Consistency then requires that $\Lambda^*$ lies below the scale at which the bound is violated. This is illustrated in figure \ref{fig:sw}.

\section{Self consistency due to string length}

There is a very simple and natural value for $\Lambda^*$ which arises due to the fact that when oscillator modes are excited the string becomes longer. This, and other interesting properties of strings with oscillators, are discussed in the review \cite{Iengo:2006gm}. The length of the string $L_{(l)}$ is expected to behave as
\be
L^2_{(l)} \sim \frac{l}{M_s^2}\;.
\label{exxstlen}
\ee
Requiring that this is less than the Hubble scale therefore implies
\be
l \lesssim \left( \frac{M_s}{H} \right)^2 \;.
\ee
Comparing with (\ref{critspin}) we see that this is precisely the critical spin value at which the Higuchi bound would be violated. This suggests that indeed when strings become of length of order the Hubble scale the application of the Higuchi bound to the Regge trajectory breaks down, and so there is no manifest inconsistency as long as $M_s > H$. 

This result is somewhat expected from the behaviour of Kerr black holes in de Sitter space. There again there is a dependence of the mass on spin $M^2 \sim l$, but there is no inconsistency with the Higuchi bound since the spin which violates it corresponds to the black hole horizon reaching the de Sitter horizon. In principle, unlike black holes, long strings could still be fitted into the de Sitter horizon. But it is natural to expect that this inevitably modifies the Regge trajectory since it is known that the maximal spin per mass for a classical string solution is just a straight rotating string. 

\section{Requiring the Regge trajectory}
\label{sec:req}

In general, if we require that the cutoff $\Lambda^*$ maintains consistency with the Higuchi bound then we have 
\be
M_{s} > \sqrt{H\Lambda^* } \;.
\label{m2la}
\ee
As discussed, taking $\Lambda^*$ to be set by the equality of string length with Hubble scale trivialises the constraint (\ref{m2la}). However, this comes at the cost of cutting off the familiar flat space string spectrum at an ultraviolet scale which is far below the Planck scale. This ultraviolet modification is a manifestation of UV/IR mixing in string theory, where the long length of strings is related to their high mass. It is known that the Regge trajectory and high spin states are a central element of the soft behaviour of string theory in the ultraviolet, or more explicitly, are responsible for modular invariance. We may therefore naturally consider the idea that decent ultraviolet behaviour may require maintaining the Regge trajectory at least until the Planck mass $M_p$. 

If we do this, then we obtain the interesting constraint
\be
\label{finalbound}
M_{s} > \sqrt{H M_p}  \;.
\ee
This constraint can in principle be considered independently of the Higuchi bound, and follows simply from using (\ref{exxstlen}) and requiring that high oscillator modes should be shorter than the de Sitter horizon scale. Nonetheless, the Higuchi bound further strengthens it in the sense that (\ref{finalbound}) would still hold even if we considered the possibility that the high oscillator modes could come from string configurations which depart from the estimate (\ref{exxstlen}), and so would fit inside the Horizon. 
 
The bound (\ref{finalbound}) can be recast into a statement about the string coupling $g_s$. Let us restrict for simplicity to compactifications to four-dimensions, but the expressions are easily generalised to any dimension. The bound is
\be
g_s > g_s^\star=\hat{R}^3 \sqrt{\frac{H}{M_p}}\;.
\label{4strcpbnd}
\ee 
Here $\hat{R}$ is the dimensionless length-scale of the extra dimensions in string units, so that in the geometric regimes we require $\hat{R} \gg 1$. 

\subsection*{String Oscillator Mode Decays}
\label{sec:strdecay}

If we consider application of the bound (\ref{finalbound}) then we should also analyse whether the required oscillator modes are sufficiently stable, so stable over the de Sitter timescale. If the state is highly unstable, then it is not clear that the Higuchi bound should even apply. Let us denote the decay rate of the spin $l$ oscillator mode as $\Gamma_{l}$, then to treat it as a stable spin-$l$ state we require
\be
\Gamma_l \ll H \;.
\label{oscdecaysta}
\ee
The physics of the decay of strings with high oscillator number is summarised in the review \cite{Iengo:2006gm}. We will focus on closed strings since they are universal. The decay rate always behaves as
\be
\Gamma_l \sim g_s^2 M_s l^r \;,
\ee
where $r$ is an ${\cal O}(1)$ power which depends on the way that the string decays. For sufficiently strongly-coupled strings the decay rate will be faster than the Hubble scale, violating the assumption (\ref{oscdecaysta}). This occurs for
\be
g_s > \frac{\hat{R}}{l^{\frac{r}{3}}} \left(\frac{H}{M_p}\right)^{\frac13}\;.
\ee
We should apply this for $l_{\mathrm{crit}}$ which yields a bound
\be
g_s > \left(\frac{H \hat{R}^3 }{M_p}\right)^{\left(\frac{1+2r}{3+2r}\right)}\;.
\label{finalgsbnd} 
\ee

The value of $r$ depends on the configuration of the string when it decays, and multiple such decay channels were studied in \cite{Iengo:2006gm}. There are permanently folded and pulsating strings, which have $r=0$, strings self-intersecting at a point, strings with end-points in contact, and strings which fold at one instant in time, which have $r=-\frac12$, and strings which self-intersect at two or more points, which have $r=-1$. The cases with $r < 0$ lead to trivial constraints, so these states are sufficiently stable in any perturbative regime. If we require the $r=0$ string configurations to be the relevant ones for the Higuchi bound, then we obtain from (\ref{finalgsbnd}) an additional constraint on the string coupling for them to be sufficiently stable. 
\newline
\subsection*{Overall bounds}

To estimate the implications of the overall bounds we note that in inflationary scenarios which lead to significant tensor modes $r \sim 0.01$ one has $\sqrt{\frac{H}{M_p}}\simeq 5 \times 10^{-3}$. Then a perturbative string regime, $g_s^\star \lesssim {\cal O}(1)$ in (\ref{4strcpbnd}), requires $\hat{R} \lesssim 4$, making control rather difficult. However, if the problematic states are string configurations which decay with $r=0$, then it is easier to simply allow them to decay too fast, which can be done in the perturbative regime with the milder constraint $\hat{R} \lesssim 30$.

\vspace{10px}
{\bf Comment:}
\noindent
The results of this work were presented during a talk at the
 conference String Phenomenology 2019:
D.  L{\"u}st,
    "Higher Spin Theories, AdS Distances and the Swampland", https://indico.cern.ch/event/782251.
After this talk, a paper 
appeared \cite{noumi}, in which also the string Regge trajectory on de Sitter space and implications for inflation were discussed.

\vspace{10px}
{\bf Acknowledgements}
\noindent
We like to thank Gia Dvali, Marvin L\"ubben and Timo Weigand for valuable discussions. We are particularly grateful to Cumrun Vafa for initial collaboration and for useful discussions.
The work of D.L. is supported by the Origins Excellence Cluster.

\bibliography{Higuchi}

\end{document}